\documentstyle[aps,epsf]{revtex}

\tighten
\newcommand{\beq}{\begin{equation}}
\newcommand{\eeq}{\end{equation}}
\newcommand{\ba}{\begin{eqnarray}}
\newcommand{\ea}{\end{eqnarray}}
\newcommand{\nn}{\nonumber}

\newcommand{\bpt}{\bm p_T^{}}
\newcommand{\bkt}{\bm k_T^{}}
\newcommand{\psibar}{\overline{\psi}}
\newcommand{\la}{\langle}
\newcommand{\ra}{\rangle}
\newcommand{\amp}[1]{\la #1 \ra}

\newcommand{\slsh}[1]{\mbox{$\not\! #1$}}
\newcommand{\bm}[1]{\bbox{#1}}
\newcommand{\simorder}{\raisebox{-4pt}{$\, \stackrel{\textstyle >}{\sim} \,$}}
\newcommand{\simordertwo}{\raisebox{-4pt}{$\, \stackrel{\textstyle <}{\sim} \,$}}

\begin{document}
 
\draft
\title{
\begin{flushright}
\begin{minipage}{4 cm}
\small
hep-ph/0102071\\
RIKEN-BNL preprint
\end{minipage}
\end{flushright}
Sudakov suppression in azimuthal spin asymmetries}

\author{Dani\"el Boer}
\address{\mbox{}\\
RIKEN-BNL Research Center\\
Brookhaven National Laboratory, Upton, New York 11973}

\maketitle
\begin{center}\today \end{center}

\begin{abstract}
It is shown that transverse momentum dependent azimuthal spin asymmetries
suffer from suppression due to Sudakov factors, in the region where
the transverse momentum is much smaller than the large energy scale $Q^2$. 
The size and $Q^2$ dependence of this suppression are studied numerically 
for two such asymmetries, both arising due to the Collins effect.  
General features are discussed of how the fall-off with $Q^2$ is affected by  
the nonperturbative Sudakov factor and by the transverse momentum weights and
angular dependences that appear in different asymmetries. For a subset of 
asymmetries the asymptotic $Q^2$ behavior is calculated analytically, 
providing an upper bound for the decrease with energy of other asymmetries. 
The effect of Sudakov factors on the transverse momentum distributions
is found to be very significant already at present-day collider energies. 
Therefore, it is essential to take into account Sudakov factors in transverse 
momentum dependent azimuthal spin asymmetries. 
\end{abstract}


\section{Introduction}

In this article we will study the effects of Sudakov factors  
in transverse 
momentum dependent azimuthal spin asymmetries, like the Collins 
effect asymmetry of Ref.\ \cite{Collins-93b}. 
We will demonstrate explicitly that such asymmetries suffer from 
suppression due to these Sudakov factors, in the region where the transverse 
momentum is much smaller than the large energy scale $Q^2$. This Sudakov 
suppression stems from
soft gluon radiation and increases with energy. It implies that tree level 
estimates of such asymmetries tend to overestimate magnitudes and
increasingly so with rising energy. In this paper, the $Q^2$ dependence of 
the suppression due to Sudakov factors will be studied numerically in two 
examples which are relevant for present-day experimental studies. In addition, 
the asymptotic $Q^2$ dependence of an important subset of asymmetries is 
calculated analytically, providing an upper bound for the decrease with energy
of other asymmetries. 
 
The first example we consider is a Collins effect driven $\cos(2 \phi)$ 
asymmetry in electron-positron annihilation into two almost back-to-back pions
\cite{Boer,BoerZ}, which in
principle can be determined from existing LEP data. This asymmetry allows for 
a determination of
the Collins effect fragmentation function, which in turn would be useful for 
the extraction 
of the transversity distribution function $h_1$. The latter can be done 
via the single spin asymmetry
in semi-inclusive Deep Inelastic Scattering (DIS) which was originally 
proposed by Collins \cite{Collins-93b} --our second example--. 
In Ref.\ \cite{Collins-93b} Collins already remarks that the Sudakov factors 
will have
the effect of diluting this single spin asymmetry due to broadening of the 
transverse momentum distribution by soft gluon emission. We will study this 
effect in a quantitative way to gain insight into the parametric dependences 
of such Sudakov suppression. 

For this we will follow the recent analysis 
\cite{Boer-00} of a helicity non-flip 
double transverse spin asymmetry in vector boson production in hadron-hadron 
scattering. That asymmetry is for instance relevant for the polarized 
proton-proton collisions to be performed at BNL-RHIC. 
In Ref.\ \cite{Boer-00} the effect of 
the Sudakov factors compared to 
the tree level asymmetry expression was numerically 
estimated. It was shown that the inclusion of Sudakov factors cause 
suppression by at least an order of magnitude compared to the tree level 
result at scales around $M_W$ or $M_Z$. Moreover, the suppression 
increases with energy approximately as $Q^{0.6}$ 
(in the studied range of roughly $10 - 100$ GeV). 
The conclusion was that the Sudakov suppression together with a kinematic
suppression (due to explicit lightcone momentum fractions that appear in the
prefactor) imply that the asymmetry will be negligible for $Z$ or $W$ 
production and is interesting only at much lower energies. 

Similarly, Sudakov
suppression will turn out to be an important issue for the actual
determination of the above mentioned $\cos(2 \phi)$ asymmetry from LEP data
(or in general at high energy $e^+ e^-$ colliders).
Due to the lack of knowledge of the nonperturbative Sudakov factor in the
case of electron-positron annihilation into almost back-to-back hadrons,
solid numerical conclusions about the size and $Q^2$ dependence of the 
suppression cannot be drawn. Nevertheless, one can draw conclusions about 
what determines the $Q^2$ dependence of the transverse momentum distribution 
of the asymmetries and about the size of the suppression for generic 
nonperturbative Sudakov factors. For that purpose  
comparisons to tree level are most instructive, since the  
evolution of the often unknown distribution and fragmentation functions
themselves does not affect the relative Sudakov suppression compared to tree
level. 

The two examples to be investigated here
--the above mentioned $\cos(2 \phi)$ asymmetry and the original Collins $\sin
(\phi)$ asymmetry in semi-inclusive DIS--, have different transverse momentum 
weights and angular dependences, producing different Sudakov suppression 
effects. General conclusions about how these properties affect the $Q^2$ 
dependence can be drawn. In general, a larger power of transverse
momentum in the weight of an asymmetry implies larger suppression. Both the
numerical calculations at realistic collider energies and the asymptotic 
behavior of asymmetries exhibit this feature. 

Before going into the details of the specific examples, 
we will first give an overview of the issues regarding 
factorization theorems for 
transverse momentum dependent cross sections, in the region where the 
transverse momentum is very small compared to the hard scale in the 
process \cite{CS81,CS85e,CSS85}. This will provide the theoretical
justification of the factorized asymmetry expressions that will be derived.

The outline of this paper is as follows. In section II we will discuss the 
essentials of factorization theorems for the Drell-Yan process \cite{CSS85} 
(relevant for
the double transverse spin asymmetry studied in Ref.\ \cite{Boer-00}), which 
are in fact completely analogous to the case of back-to-back jets in 
electron-positron annihilation \cite{CS81,CS85e} and semi-inclusive DIS 
(SIDIS). Following the approach of Ref.\ \cite{Boer-00}, we will then study 
the $\cos (2 \phi)$ asymmetry in 
electron-positron annihilation (section III) and the Collins $\sin
(\phi)$ asymmetry in semi-inclusive DIS (section IV). 
In section V we will discuss the asymptotic behavior of the Sudakov
suppression for certain subsets of asymmetries.

\section{Factorization and transverse momentum}

For definiteness, we will consider here the Drell-Yan process ($H_1 + H_2
\rightarrow \ell + \bar\ell + X$) and discuss what is known about the
factorization of this process. First of all, it is well-known that a local
operator product expansion (OPE) cannot be applied, since the process 
receives relevant contributions off the light cone \cite{Jaffe-96}, i.e.\ 
from matrix elements like
\beq
\sum_X \langle P_1, P_2| J_\mu ({\xi})| X \rangle
\, \langle X | J_\nu (0)| P_1, P_2 \rangle \qquad \text{with} \;\; {\xi^2
\neq 0}.
\eeq
Therefore, one usually considers nonlocal operators and what is called 
a ``working redefinition of twist'' \cite{Jaffe-96}, which means the lowest 
value of $t$ in $1/Q^{t-2}$ at which a function can contribute to the cross
section ($Q$ is the invariant mass of the lepton pair).  
The nonlocal operators appear in the leading twist factorization theorem 
\beq
\frac{d\sigma}{d\Omega dx_1 dx_2} = \sum_{a,b} \int_{x_1}^1 dx\,
\int_{x_2}^{1} d\bar{x}\, {\Phi^a(x)}\, H^{ab}(x,\bar{x}; Q^2) 
\, {\overline \Phi}{}^b(\bar{x}),
\label{collinearfact}
\eeq
where $a, b$ are flavor indices and 
$H^{ab}(x,\bar{x}; Q^2)$ is the hard (partonic) part of the scattering, which
is a function of the hard scale $Q^2$ and the lightcone momentum fractions
$x,\bar{x}$ only. The
correlation functions $\Phi$ ($\overline \Phi{}^a = \Phi^{\bar{a}}$) 
describe the soft (nonperturbative) physics and 
are given by nonlocal operator matrix elements
\ba
{\Phi_{ij} (x)} 
&\equiv& \int \frac{d \lambda}{2\pi}e^{i\lambda x}\amp{P,S|\,
\psibar_j (0) \, {{\cal L}^+ [0,\lambda n_-]} 
\, \psi_i(\lambda n_-) | P,S},
\ea
where the path-ordered exponential, 
\beq
{{\cal L}^+ [0,\lambda n_-]} = {\cal P} 
\exp \left(-ig\int_{0}^{\lambda} d \eta \, A^+(\eta n_-) 
\right),
\eeq
renders the matrix element color gauge invariant. The hadronic state $| P,S
\rangle $ is determined by the hadron momentum $P$ and spin vector $S$. 
Also, we have suppressed the factorization scale dependence in 
Eq.\ (\ref{collinearfact}).

Secondly, if one considers the cross section differential in the transverse 
momentum $\bm{q}_T^{}$ of the lepton pair, then for the case of 
$\bm{q}_T^2 \equiv Q_T^2 \ll Q^2$ collinear factorization, like in Eq.\
(\ref{collinearfact}), does not hold. Instead, the leading twist factorization 
takes the form \cite{CS81,CS85e,CSS85,CSS83,CSS-PLB84}
\ba
\frac{d\sigma}{d\Omega dx_1 dx_2 d^2 \bm{q}_T}& =& 
\sum_{a,b} \int_{x_1}^1 dx\, \int_{x_2}^{1} d\bar{x}\, 
\int d^2 \bm{k}_T \, d^2 \bm{p}_T \, \int \frac{d^2 \bm{b}}{(2\pi)^2} \, 
e^{-i \bm{b} \cdot (\bm{p}_T^{}+\bm{k}_T^{}-\bm{q}_T^{})} \nn \\[2 mm] 
&& \times \Phi^a(x,\bm{k}_T^{})\,
H^{ab}(x,\bar{x};Q) \, e^{-S(\bm{b},Q)} \, 
\overline \Phi{}^b(\bar{x},\bm{p}_T^{}) 
+ Y(x_1,x_2,Q,Q_T^{}). 
\label{facttheorem}
\ea
The correlation function $\Phi^a(x,\bm{k}_T^{})$ is now also a function of
transverse momentum, the factor $e^{-S(\bm{b},Q)}$ is the so-called Sudakov 
form factor and the factor $Y(x_1,x_2,Q,Q_T)$ becomes important only when 
$Q_T \sim Q$. Here we use a different notation than in e.g.\  
Refs.\ \cite{CS81,CS85e,CSS85,CSS83,CSS-PLB84}, but to make a connection with 
for instance the notation of Ref.\ \cite{CS81} we note that
$\Phi(x,\bm{k}_T^{})$ corresponds to the Fourier 
transform of $\hat{\cal P}_{A/i}(x_A,\bm{b}) \slsh{P}_A/2$ and 
the $e^{-S}$ terms correspond to each other. 

The factor $Y(x_1,x_2,Q,Q_T)$ is present to retrieve collinear 
factorization when $Q_T$ is a hard scale itself, 
i.e.\ $Q_T \sim Q$. We will neglect the $Y$ term from here on, 
since we will only be interested in the region of $Q_T^2 \ll Q^2$. We will 
now comment on the other terms in the above factorization theorem and explain 
the additional restrictions we will impose in the limit $Q_T^2 \ll Q^2$. 

The factor $e^{-S(\bm{b},Q)}$ is the so-called Sudakov form factor, which 
arises due to
exponentiation of soft gluon contributions. This is in contrast to 
inclusive cross sections, like Eq.\ (\ref{collinearfact}), 
in which there is a cancellation of soft gluon contributions. 
At values $b^2=\bm{b}^2 \ll 1/\Lambda^2$, the Sudakov form factor is
perturbatively calculable and is of the form 
\beq
S(b,Q)=\int_{b_0^2/b^2}^{Q^2} \frac{d \mu^2}{\mu^2} \left[ 
A(\alpha_s (\mu)) \, \ln \frac{Q^2}{{\mu}^2} + 
B (\alpha_s (\mu)) \right],
\label{sud}
\eeq
where $b_0=2\exp(-\gamma_E) \approx 1.123$ (we choose the usual 
constants $C_1=b_0, C_2=1$). 
One can expand the functions $A$ and $B$ in $\alpha_s$ and the first few 
coefficients are known, e.g.\ \cite{Davies-Stirling,Weber}. In order to
obtain a first estimate of the effect of including the Sudakov factor we will 
consider only the leading contribution, i.e.\ take into account only the 
first term in the expansion of $A$: $A^{(1)} =  C_F/\pi$, which is the same
for unpolarized as well as polarized scattering. This leads to the
expression \cite{Frixione}
\beq
S(b,Q)=-\frac{16}{33-2n_f} \left[ \log\left(\frac{b^2 Q^2}{b_0^2}\right)+
\log\left(\frac{Q^2}{\Lambda^2}\right)\; \log\left[1- \frac{\log\left(b^2 
Q^2/b_0^2\right)}{\log\left(Q^2/\Lambda^2\right)} \right]\right].
\label{sudactual}
\eeq
We will take for the number of flavors $n_f =5$ and also $\Lambda_{QCD}=200 \,
\text{MeV}$. 

The correlation function $\Phi^a(x,\bm{k}_T^{})$ is now defined as
\cite{CSS83,Boer-Mulders-00}
\beq
\Phi(x,\bm{k}_T^{}) \equiv  \left. \int dk^-\ \left[
\int \frac{d^4x}{(2\pi)^4}\, e^{ik\cdot x}\, \langle P, S \vert \overline
\psi(0) \, {\cal L}^+ [0,-\infty] \, {\cal L}^+ 
[-\infty,x]
\, \psi(x) \vert P, S\rangle \right] \right|_{k^+ = x P^+, \bkt},
\eeq
with for each colored field a path-ordered exponential like  
\begin{equation}
{\cal L}^+[-\infty,x] = \left. {\cal P} \exp \left(- ig \int_{-\infty}^{x^-} 
dy^- \, A^+(y) \right)\right|_{y^+ = x^+, \bm{y}_T=\bm{x}_T}.
\label{linkDY}
\end{equation}
The correlation function $\Phi^a(x,\bm{k}_T^{})$ also has a factorization
scale dependence ($\mu=b_0/b$), which is linked to that of the hard 
scattering part $H$. 

The correlation function $\Phi^a(x,\bm{k}_T^{})$ contains the nonperturbative
dependence on the transverse momentum, which cannot be calculated
perturbatively and should be fitted to experiment. 
In for example Refs.\ \cite{CS81,CS85e,CSS85} it is
explained how these functions can be replaced by the ordinary
parton distribution functions $\Phi^a(x)$, by introduction of a $b$-regulator,
e.g.\ the usual cut-off $b_{\max}$, via 
$b_*=b/\sqrt{1+b^2/b_{\max}^2}$, and by introducing a nonperturbative Sudakov
factor $S_{NP}$. Schematically, this proceeds as follows. The first term in 
Eq.\ (\ref{facttheorem}) can be written as an overall $\bm{b}$ 
integration of an integrand called $\tilde{W}(b)$, 
\beq
\frac{d\sigma}{d\Omega dx_1 dx_2 {d^2 \bm{q}_T}} = \int_{x_1}^1 dx\,
\int_{x_2}^{1} d\bar{x}\, \int 
\frac{d^2 \bm{b}}{(2\pi)^2} e^{i {\bm{b} \cdot
{\bm{q}_T^{}}}} {\tilde{W}(b)} + Y.
\eeq
This function $\tilde{W}(b)$ can be trivially 
rewritten as $\tilde{W}(b_*) \times (\tilde{W}(b)/\tilde{W}(b_*))$. 
The first term ($\tilde{W}(b_*)$) can be
calculated within perturbation theory since $\alpha_s(b_0/b_*)$ is always
small (we will take the usual $b_{\max}=0.5 \;
\text{GeV}^{-1}$, hence $\alpha_s(b_0/b_* \simorder 2) \simordertwo 0.3$) 
and the second term can be shown to be of the form 
\beq
\frac{\tilde{W}(b)}{\tilde{W}(b_*)} = 
\exp\left(-\ln(Q^2/Q_0^2) g_Q(b) - g_A(x,b) - g_B(\bar x, b)\right)
\equiv \exp\left(-S_{NP}(b)\right).
\label{genNPform}
\eeq
The functions $g$ are not calculable in perturbation theory and need to be 
fitted to experiment. The functions $g_A(x,b)$ and
$g_B(\bar x,b)$ parameterize the intrinsic transverse momentum of the
functions $\tilde{\Phi}{}^a(x,b)$ and
$\tilde{\overline{\Phi}}{}^b(\bar{x},b)$ 
(in the polarized case $\tilde{\Phi}{}^a$ is
actually a function of $\bm{b}$, see further comments below). 
It will be important later on that the functions $g_A$ and $g_B$  
are independent of the scale $Q$. Finally, a lightcone expansion allows to 
express $\tilde{\Phi}^a(x,b_*)$ in terms of $\Phi^a(x)$.
 
We will consider $\tilde{\Phi}^a(x,\bm{b}_*) \tilde{\overline{\Phi}}{}^b(\bar
x,\bm{b}_*) H(x,\bar x;Q)$ only to lowest order in $\alpha_s$. 
The reason is that the perturbative tail of 
$\tilde{\Phi}^a(x,\bm{b}_*)$ is not known 
for the Collins function. Because of the unknown 
size of the distribution and/or fragmentation functions appearing in the 
asymmetries, the magnitude of the asymmetries cannot be estimated. 
Therefore, here we will be interested in the effect of soft gluons on the 
transverse momentum dependence of the asymmetries, which determines 
the $Q^2$ dependence of the $Q_T$ dependence. Although the hard scattering 
part is restricted to tree 
level, one can draw conclusions on the size of the effect of Sudakov factors 
by comparison to the tree level result, because both results will receive the 
same corrections to the hard part. Corrections to the hard
scattering part can be easily
inserted with hindsight (for a one-loop
example cf.\ Ref.\ \cite{CS81}, pages 432 and 433). 
As said we are interested in the $Q^2$ evolution of the 
$Q_T$ dependence of the asymmetries. This $Q^2$ dependence arises in a
nontrivial way via Bessel functions, such that a numerical study is 
warranted.  

In Refs.\ \cite{CS81,CS85e,CSS85} unpolarized scattering is
considered, where only dependence on the length of $\bm{b}$ appears. 
Here we will
consider polarized scattering and the above sketched procedure of Refs.\
\cite{CS81,CS85e,CSS85} to factor a nonperturbative part from
$\tilde{W}(b)$, can also be applied after $\Phi^a(x,\bm{k}_T^{})$ is
parameterized in terms of functions of $\bm{k}_T^{2}$. Explicit factors of
$\bm{k}_T^{} \! \cdot \! \bm{S}_T^{}$ or $\slsh{\bm{k}_T^{}}$ can be included 
in the hard scattering part $H$. 

The above factorization theorem for the Drell-Yan process provides the 
justification of the factorized expression for
the double transverse spin asymmetry studied in Ref.\ \cite{Boer-00}.
In Ref.\ \cite{CS81,CS85e} an 
analogous factorization theorem had been discussed earlier for
back-to-back jets in electron-positron annihilation, which comes down to a
replacement of $\Phi(x,\bm{k}_T^{}) \to \Delta(z,\bm{k}_T^{})$ --the 
fragmentation correlation function--. 
In Ref.\ \cite{Collins-93b} it is argued 
that a similar factorization theorem holds for semi-inclusive DIS. Therefore,
factorization theorems analogous to the one discussed above (differing only by
obvious replacements), provide the theoretical 
foundation for the examples to be considered below. 

It is interesting to see what happens if one takes tree level everywhere in
the above factorized expression Eq.\ (\ref{facttheorem}) (ignoring the $Y$
term). At tree level one finds the reduction 
\beq
\int \frac{d^2 \bm{b}}{(2\pi)^2} \, 
e^{-i \bm{b} \cdot
(\bm{p}_T^{}+\bm{k}_T^{}-\bm{q}_T^{})} \, e^{-S(\bm{b})} \to {\delta^2
( \bm{p}_T^{}+\bm{k}_T^{}-\bm{q}_T^{})},
\eeq
such that one arrives at the expression used by Ralston and Soper for the 
hadron tensor \cite{Ralst-S-79,Tangerman-Mulders-95a,Boer-00}
\beq
{\cal W}^{\mu\nu}=\frac{1}{3} \int d^2\bm{p}_T^{} d^2 
\bm{k}_T^{}\, \delta^2(\bm{p}_T^{}+
\bm{k}_T^{}-\bm{q}_T^{})\, \left.
\text{Tr}\left( \Phi (x_1,\bpt) \, 
V_1^\mu \, \overline \Phi  
(x_2,\bkt) \, V_2^\nu \right) \right|_{p^+, \, k^-}
+ \left(\begin{array}{c} 
q\leftrightarrow -q \\ \mu \leftrightarrow \nu
\end{array} \right).
\label{hadrontensor}
\eeq
The vertex $V_i^\mu$ can be either the photon, $Z$ or $W$ boson vertex.
In the above unpolarized case $\Phi (x,\bpt)$ is parameterized as 
$\frac{1}{2} {f_1(x,\bm{p}_T^2)} \mbox{$\not\! P\,$}$, where $f_1$ is the
parton momentum distribution function. 
Collins \cite{Collins-93a} has shown that polarization 
does not affect the factorization theorems and therefore, we conclude that 
the tree level formalism of Ralston and Soper Eq.\ (\ref{hadrontensor}) is in
accordance with the factorization theorem Eq.\ (\ref{facttheorem}) also in 
the case when polarization is taken into account. 
Clearly, Eq.\ (\ref{hadrontensor}) is only applicable for $Q_T$ values
of the order of the intrinsic transverse momentum. To extend the 
formalism of Ralston and Soper to
larger values of $Q_T$, but still under the restriction of $Q_T^2 \ll Q^2$, one
can use the above factorized expression Eq.\ (\ref{facttheorem}) (without the
$Y$ term) beyond tree level. 

\section{Unpolarized asymmetry in electron-positron annihilation}

In this section we investigate a $\cos(2 \phi)$ asymmetry in the process of 
electron-positron annihilation into two almost back-to-back pions
$e^+ e^- \to \pi^+ \pi^- X$. This process is very similar to the
Drell-Yan process, so we can employ the same expression as given in Eq.\
(\ref{facttheorem}), but with the correlation functions $\Phi$, replaced by  
the fragmentation correlation function
\cite{Coll-S-82}
\beq
\Delta_{ij}(k) =  \sum_X \int \frac{d^4z}{(2\pi)^4}\ e^{ik\cdot z}
\langle 0 \vert \psi_i(x) \, {\cal L}^- [x,\infty] \vert P,S; X \rangle 
\langle P, S; X \vert {\cal L}^- [\infty,0] \, \overline \psi_j(0) \vert 0 
\rangle,  
\eeq 
which is used to define transverse momentum dependent fragmentation functions.

The asymmetry 
requires observing the transverse momentum of the vector boson compared to the
two pions. The tree level 
asymmetry expression for the $\cos(2 \phi)$ asymmetry in the process 
$e^+ e^- \to \pi^+ \pi^- X$ was discussed in Refs.\ 
\cite{Boer,BoerZ}, to which we refer for details. 
Here we shortly repeat the essentials. 
The asymmetry depends on the fragmentation function $H_1^\perp$ 
associated with the Collins effect \cite{Collins-93b}. 
The definition of the Collins function is given by
\cite{Mulders-Tangerman-96}
\beq
\Delta(z,\bkt)= \frac{M}{4P^-}\,\Biggl\{
D_1(z ,\bm{k}_T^2)\, \frac{\slsh{\! P}}{M} 
+ H_{1}^\perp(z,\bm{k}_T^2)\,\frac{\sigma_{\mu \nu} k_T^\mu P^\nu}{M^2}
\biggr\},
\eeq
where we have only displayed the fragmentation functions that can be present 
for unpolarized hadron production ($D_1$ is the ordinary unpolarized
fragmentation function). The $\cos(2 \phi)$ asymmetry is an
azimuthal spin asymmetry in the sense that the asymmetry arises due to the
correlation of the transverse spin states of the quark-antiquark pair. On
average the quark and antiquark will not be transversely polarized, but for
each particular event the transverse spin need not be zero and moreover, 
the spin states of the quark and antiquark are exactly
correlated. Subsequently, the directions of the 
produced pions are correlated due to the Collins effect and this correlation 
does not average out after 
summing over all quark polarization states. The Collins effect correlates the
azimuthal angle of the transverse spin of a fragmenting quark with that  
of the transverse momentum of the produced hadron (both taken around 
the quark momentum), 
via a $\sin (\phi)$ distribution of their difference angle $\phi$.  

One finds for the leading order unpolarized cross section, taking into
account both photon and $Z$-boson contributions \cite{Boer,BoerZ}, 
\ba
\lefteqn{
\frac{d\sigma (e^+e^-\to h_1h_2X)}{d\Omega dz_1 dz_2 
d^2{\bm q_T^{}}} =
\frac{3\alpha^2}{Q^2}\;z_1^2z_2^2\;\sum_{a,\bar a} \;\Bigg\{ 
          K_1^a(y)\;{\cal F}\left[D_1\overline D_1\right]}
\nonumber\\ && 
       \qquad + \left[K_3^a(y)\cos(2\phi_1)+K_4^a(y)\sin(2\phi_1)\right]\;
             {\cal F}\left[\left(2\,\bm{\hat h}\!\cdot \!
\bm p_T^{}\,\,\bm{\hat h}\!\cdot \! \bm k_T^{}\,
                    -\,\bm p_T^{}\!\cdot \! \bm k_T^{}\,\right)
                    \frac{H_1^{\perp}\overline H_1^{\perp}}{M_1M_2}\right]
\Bigg\}.
\label{LO-OOO}
\ea
The convolutions ${\cal F}$ will be dicussed below (Eqs.\
(\ref{conv}) and (\ref{conv2})), but first we will give the definition of 
the functions $K_i^a(y)$ appearing in this expression and comment on the 
frame in which this cross section is expressed. As before, $a$ is the flavor 
index; the functions $K_i^a(y)$ are defined as  
\ba
K_1^a(y) &=& A(y)\;
\left[ e_a^2+ 2 g_V^l e_a g_V^a \chi_1 + c_1^l c_1^a \chi_2
\right] 
- \frac{C(y)}{2} \; \left[ 2 g_A^l e_a g_A^a \chi_1 + c_3^l c_3^a \chi_2 
\right],\\
K_3^a(y)&=& B(y)\; \left[ e_a^2+ 2 g_V^l e_a g_V^a \chi_1 + c_1^l c_2^a 
\chi_2 \right], \\ 
K_4^a(y)&=& B(y)\; \left[ 2 g_V^l e_a g_A^a \chi_3 \right],
\ea
which contain the combinations of the couplings
\begin{eqnarray} 
c_1^j &=&\left(g_V^j{}^2 + g_A^j{}^2 \right),\\
c_2^j &=&\left(g_V^j{}^2 - g_A^j{}^2 \right),\\
c_3^j &=&2 g_V^j g_A^j,
\end{eqnarray}
where $g_V$ and $g_A$ are the vector and axial-vector couplings to the $Z$
boson. The propagator factors are given by 
\ba
\chi_1 &=& \frac{1}{\sin^2 (2 \theta_W)} \, \frac{Q^2
(Q^2-M_Z^2)}{(Q^2-M_Z^2)^2 + \Gamma_Z^2 M_Z^2},\\
\chi_2 &=& \frac{1}{\sin^2 (2 \theta_W)} \, \frac{Q^2}{Q^2-M_Z^2} \chi_1,\\
\chi_3 &=& \frac{-\Gamma_Z M_Z}{Q^2-M_Z^2} \chi_1.
\ea

The above cross section is expressed in the following frame. A normalized 
timelike vector $\hat{t}$ is defined by $q$ (the vector boson momentum) and
a normalized spacelike vector $\hat{z}$ is defined by $\tilde P^\mu$ = 
$P^\mu - (P\cdot q/q^2)\,q^\mu$ for one of the outgoing momenta, for which we 
choose $P_2$ ($P_i$ are the hadron momenta),
\begin{eqnarray}
\hat t^\mu& \equiv & \frac{q^\mu}{Q},\\
\hat z^\mu & \equiv &
\frac{Q}{P_2\cdot q}\,\tilde P^\mu_2
\ =\   2\,\frac{P_2^\mu}{z_2 Q} - \frac{q^\mu}{Q},
\end{eqnarray}
The azimuthal 
angles lie inside the plane orthogonal to $\hat{t}$ and $\hat{z}$. 
In particular, $\phi^\ell$ gives the 
orientation of $\hat l_\perp^\mu \equiv \left( g^{\mu \nu}-\hat t^{ 
\mu} \hat t^{\nu } + \hat z^{
\mu} \hat z^{\nu } \right) l_\nu$, the perpendicular part of the lepton
momentum $l$. 

In the cross sections we also 
encounter the following functions of $y=l^-/q^-$, which in the
lepton center of mass frame equals $y=(1 + \cos \theta_2)/2$, where 
$\theta_2$ is the angle of $\hat z$ with respect to the momentum of the 
incoming lepton $l$:
\ba
A(y) &=& \left(\frac{1}{2} -y+y^2\right) \stackrel{cm}{=} 
\frac{1}{4} \left( 1 + \cos^2\theta_2 \right)
, \\
B(y) &=& y\,(1-y) \stackrel{cm}{=}\frac{1}{4} \sin^2 \theta_2 
,\\[2 mm]
C(y) &=& (1-2y) \stackrel{cm}{=} -\cos\theta_2.
\ea

\begin{figure}[htb]
\begin{center}
\leavevmode \epsfxsize=10cm \epsfbox{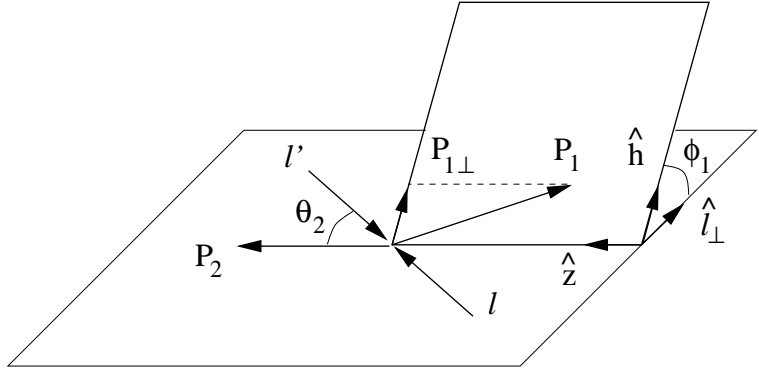}
\vspace{0.2 cm}
\caption{\label{fig:kinann} Kinematics of the annihilation process in
the lepton center of mass frame for a back-to-back jet situation.
$P_2$ is the momentum of a hadron in one jet, $P_1$ is
the momentum of a hadron in the opposite jet.}
\vspace{-2 mm}
\end{center}
\end{figure}

Since we have chosen $P_2$
to define the longitudinal direction, the momentum $P_1$
can be used to express the directions orthogonal to 
$\hat t$ and $\hat z$. One obtains $P_{1\perp}^\mu$ = 
$g_\perp^{\mu\nu}\,P_{1\nu}$ (see Fig.\ \ref{fig:kinann}),
\begin{equation}
P_{1\perp}^\mu = - z_1\,q_T^\mu = z_1\, Q_T\, \hat h^\mu,
\end{equation}
where we define 
the normalized vector $\hat h^\mu$ = $P_{1\perp}^\mu/\vert \bm P_{1\perp}
\vert$. 
The angle $\phi_1$ is between $\bm{\hat h}$ and $\bm{\hat l}_\perp$.

The asymmetry $A(\bm{q}_T^{})$ at the $Z$ mass 
(ignoring the photon and interference
contributions, which can be easily included) is now defined by:
\beq
\frac{d\sigma (e^+e^-\to h_1h_2X)}{d\Omega dz_1 dz_2 d^2{\bm q_T^{}}} \propto 
\left\{ 1 + \cos(2\phi_1) A(\bm{q}_T^{}) \right\},
\eeq
with 
\ba
A(\bm{q}_T^{}) &\equiv &\frac{
\sum_{a}\;c_1^e\,c_2^a\;B(y)\;{\cal F}\left[\left(2\,\bm{q}_T^{}\!\cdot \!
\bm p_T^{}\,\,\bm{q}_T^{}\!\cdot \! \bm k_T^{}\,
                    -\,\bm{q}_T^{2}\,\bm p_T^{}\!\cdot \! \bm k_T^{}\,\right)
                    H_1^{\perp}\overline H{}_1^{\perp}
\right]}{Q_T^2 {M_1M_2}\sum_{b}\; \left(c_1^e\,c_1^b\;A(y)-{\scriptstyle 
\frac{1}{2}}
\,c_3^e\,c_3^b\,C(y)\right)\; {\cal F}\left[D_1\overline D_1\right]}.
\label{AQT}
\ea
A confirmation of this asymmetry would confirm the Collins effect without the 
need of polarization. To avoid repeating irrelevant factors, we 
will first focus on the numerator and denominator of the term 
\beq
\kappa =  \frac{
{\cal F}\left[\left(2\,\bm{q}_T^{}\!\cdot \!
\bm p_T^{}\,\,\bm{q}_T^{}\!\cdot \! \bm k_T^{}\,
                    -\,\bm{q}_T^{2}\,\bm p_T^{}\!\cdot \! \bm k_T^{}\,\right)
                  H_1^{\perp}\overline H_1^{\perp}\right]
}{Q_T^2\; {M_1M_2}\; {\cal F}\left[D_1\overline D_1\right]}.
\label{kappa1}
\eeq

In Refs.\ \cite{Boer,BoerZ} tree level was considered, which means the
expressions are valid in the region 
where the observed transverse momentum is very small compared to the hard 
scale(s), applicable only in the region of
intrinsic transverse momentum. 
Eq.\ (\ref{kappa1}) is a tree level expression if one uses the convolution 
notation 
\begin{equation} 
{\cal F}\left[D\overline D\, \right]\equiv \;
\int d^2\bm p_T^{}\; d^2\bm k_T^{}\;
\delta^2 (\bm p_T^{}+\bm k_T^{}-\bm 
q_T^{})  D^a(z_{1},z_{1}^2\bm{p}_T^2) 
\overline D{}^a(z_{2},z_{2}^2\bm{k}_T^2),
\label{conv}
\end{equation}
which only involves intrinsic transverse momenta. 
In order to go beyond this region, we include the Sudakov factor 
arising from resummed perturbative corrections to the transverse momentum
distribution, by considering the approach of 
Ref.\ \cite{CS81,CS85e}. This will extend the range of applicability 
from the region of intrinsic transverse
momentum to the region of moderate $\bm{q}_T^{}$ values (still under the 
restriction that $\bm{q}_T^2 \ll Q^2$). This is outlined for the general case 
in the previous section; here we will apply it to the $\cos(2\phi)$ 
asymmetry. 

Resummation of soft gluons into
Sudakov form factors (see e.g.\ \cite{Col-89}) results in a replacement in 
Eq.\ (\ref{conv}) of 
\beq
\delta^2(\bm{p}_T^{}+\bm{k}_T^{}-\bm{q}_T^{})\to \int 
\frac{d^2 \bm{b}}{(2\pi)^2} \, e^{-i \bm{b} \cdot
(\bm{p}_T^{}+\bm{k}_T^{}-\bm{q}_T^{})} \, e^{-S(b)}, 
\label{sudakovreplacement}
\eeq 
leading to (suppressing the flavor indices)  
\ba 
{\cal F}\left[D\overline D\, \right]& \equiv & \; \int 
\frac{d^2 \bm{b}}{(2\pi)^2} \, e^{i \bm{b} \cdot \bm{q}_T^{}} 
\, e^{-S(b)}\, \tilde{D}(z_{1},b) \, 
\tilde{\overline D}(z_{2},b)\nn \\[2 mm]
&=& \frac{1}{2 \pi} \int_0^\infty db \, b \, J_0(b Q_T)\, e^{-S(b)}
\tilde{D}(z_{1},b) \, \tilde{\overline D}(z_{2},b).
\label{conv2}
\ea
The function $\tilde{D}$ denotes the Fourier transform of $D$. 
In order to compute the above expression, we will assume a 
Gaussian transverse momentum dependence for $D_1(z, z^2\bm{k}_T^2)$:
\beq
D_1(z, z^2\bm{k}_T^2) = D_1(z) \; R_u^2 
\exp(-R_u^2 \bm{k}_T^2) / \pi z^2 \equiv 
D_1(z) \; {\cal G}(|\bm{k}_T^{} |; R_u)/ z^2. 
\label{Gauss}
\eeq
For details see Ref.\ \cite{Mulders-Tangerman-96}. 
Taking the Fourier transform of Eq.\ (\ref{Gauss}) yields
\beq
\tilde{D}{}_{1}^{}(z,b^2) 
= D_{1}(z) \; \exp\left(-\frac{b^2}{4 R_u^2} \right)/z^2.
\label{GaussFT}
\eeq  

The numerator in Eq.\ (\ref{kappa1}), 
\ba 
\lefteqn{{\cal F}\left[\left(2\,\bm{q}_T^{}\!\cdot \!
\bm p_T^{}\,\,\bm{q}_T^{}\!\cdot \! \bm k_T^{}\,
                    -\,\bm{q}_T^{2}\,\bm p_T^{}\!\cdot \! \bm k_T^{}\,\right)
\; D\overline D\, \right] \equiv  \; \int 
\frac{d^2 \bm{b}}{(2\pi)^2} \, e^{i \bm{b} \cdot \bm{q}_T^{}} 
\, e^{-S(b)}}\nn \\
&& \times \int d^2\bm p_T^{}\; d^2\bm k_T^{}\; \left(2\,\bm{q}_T^{}\!\cdot \!
\bm p_T^{}\,\,\bm{q}_T^{}\!\cdot \! \bm k_T^{}\,
                    -\,\bm{q}_T^{2}\,\bm p_T^{}\!\cdot \! \bm k_T^{}\,\right)
\; e^{-i \bm{b} \cdot (\bm{p}_T^{}+\bm{k}_T^{})}\; D(z_{1},z_{1}^2\bm{p}_T^2) 
\overline D(z_{2},z_{2}^2\bm{k}_T^2),
\label{bconv}
\ea
cannot be treated exactly like the denominator. A model for the transverse
momentum dependence of the 
function $H_1^\perp$ is needed. In order to get a first estimate, or rather
an upper bound for the asymmetry, one might be inclined to assume the
maximally allowed function, by saturating the bound satisfied by $H_1^\perp$
\cite{Boglione-Leader} 
\beq
|\bm{k}_T^{}| \; \left| H_1^{\perp}(z,|\bm{k}_T^{}|) \right| \leq z \; M_h \; 
D_1(z,|\bm{k}_T^{}|), 
\label{bound}
\eeq
producing a $1/|\bm{k}_T^{}|$ behavior of $H_1^{\perp}(z,\bm{k}_T^{})$.
However, this is not consistent with the fact that the Collins effect should
vanish in the limit $\bm{k}_T^{} \to 0$.  

A model by Collins \cite{Collins-93b}
suggests the following transverse momentum dependence. 
Collins' parameterization for the fragmentation function 
$H_1^\perp$ is (note that Collins uses the function 
$\Delta \hat{D}_{H/a} \sim \epsilon_T^{ij} s_{Ti} k_{T j} \, H_1^\perp$, where
$s_T$ is the transverse spin of the fragmenting quark)
\beq
\frac{H_1^\perp(z, \bm{k}_T^2)}{D_1(z,\bm{k}_T^2)} = 
\frac{2 M_C M_h}{\bm{k}_T^2+ M_C^2} \, 
\text{Im}\left[A^*(k^2) B(k^2) \right] \,
\frac{(1-z)}{z},
\eeq
where $M_h$ is the mass of the produced hadron and $M_C$ is the 
quark mass that appears in a dressed fermion propagator 
$i(A(k^2) \slsh{k} + B(k^2) M_C) /(k^2 -M_C^2)$, the functions $A$ and $B$ are
unity at $k^2=M_C^2$. But for the present purpose, the additional fall-off
with $1/\bm{k}_T^2$ on top of the Gaussian fall-off is not needed. We will
restrict to a Gaussian fall-off (with unknown magnitude) and assume the 
simple form for $H_1^{\perp a}(z,z^2\bm{k}_T^2) = H_1^{\perp a}(z) \; 
{\cal G}(|\bm{k}_T^{} |; R)/ z^2$. Here one should take the radius $R$ to be 
larger than $R_u$ of the unpolarized function $D_1$, such as to satisfy the
bound Eq.\ (\ref{bound}) for all $|\bm{k}_T^{} |$.
Also, we will assume that the fragmentation functions for both 
hadrons are Gaussians of equal width, i.e.\ we take $R_1=R_2=R$ and
$R_{u1}=R_{u2}=R_u$ and also $M_1=M_2=M$. 

One finds
\ba   
\lefteqn{
\int d^2\bm p_T^{}\; d^2\bm k_T^{}\; \left(2\,\bm{q}_T^{}\!\cdot \!
\bm p_T^{}\,\,\bm{q}_T^{}\!\cdot \! \bm k_T^{}\,
                    -\,\bm{q}_T^{2}\,\bm p_T^{}\!\cdot \! \bm k_T^{}\,\right)
\; e^{-i \bm{b} \cdot (\bm{p}_T^{}+\bm{k}_T^{})}\; {\cal
G}(\bm{p}_T^2;R) \;  
{\cal G}(\bm{k}_T^2;R) 
= } \nn \\
&& \mbox{} \hspace{6 cm} -\frac{1}{4 R^4} \; \left[ 2(\bm{q}_T^{} \cdot \bm{b})^2 - \bm{q}_T^2
\bm{b}^2 \right]\; \exp\left(-\frac{b^2}{2 R^2}\right),
\ea 
which after application to Eq.\ (\ref{bconv}) yields 
\beq
{\cal F}\left[\left(2\,\bm{q}_T^{}\!\cdot \!
\bm p_T^{}\,\,\bm{q}_T^{}\!\cdot \! \bm k_T^{}\,
                    -\,\bm{q}_T^{2}\,\bm p_T^{}\!\cdot \! \bm k_T^{}\,\right)
\; D\overline D\, \right]= 
\int 
\frac{d^2 \bm{b}}{(2\pi)^2} \, e^{i \bm{b} \cdot \bm{q}_T^{}} \,
(-\frac{1}{4 R^4}) \; \left[ 2(\bm{q}_T^{} \cdot \bm{b})^2 - \bm{q}_T^2
\bm{b}^2 \right]\; e^{-S(b)}\tilde{D}(z_{1},b) \, 
\tilde{\overline D}(z_{2},b). 
\eeq
It is important to note that the factor $1/R^4$ stems from 
the intrinsic transverse momentum of the functions $H_1^\perp(z,\bm{k}_T^{})$
and therefore, is not dependent on the scale $Q$ (see discussion after Eq.\
(\ref{genNPform})).

In analogy to $A_{TT}^V$ \cite{Boer-00}, 
Eq.\ (\ref{kappa1}) can be transformed into (keeping
in mind that in the end the summation over $a, b$ in numerator and denominator
should be performed separately)
\beq
\kappa = \frac{H_1^{\perp a}(z_1) \; \overline H{}_1^{\perp a}(z_2)}{4 M^4 R^4
\; D_1^b(z_1) \; \overline D{}_1^b(z_2)} \; {\cal A}(Q_T) = \frac{H_1^{\perp
(1) a}(z_1) \; \overline H{}_1^{\perp (1) a}(z_2)}{D_1^b(z_1) \; 
\overline D{}_1^b(z_2)} \; {\cal A}(Q_T), 
\label{kappa3}
\eeq
where for future reference we have also expressed the asymmetry in terms of 
the function ${H_1^{\perp (1)}(z)} = z^2 \int d^2 \bkt \, \bm{k}_T^2/(2M^2)\, 
H_1^\perp (z,z^2\, \bm{k}_T^2)$ which occurs frequently in $Q_T$-weighted cross
sections (e.g.\ \cite{Boer}); furthermore, 
\beq
{\cal A}(Q_T) \equiv M^2 \, \frac{
\int_0^\infty db \, b^3 \, J_2(b Q_T) \,  
\exp\left({-S(b_*)}\, 
{-S_{NP}(b)}\right)}{\int_0^\infty db \, b \, J_0(b Q_T)\, 
\exp\left({-S(b_*)}\, {-S_{NP}(b)} \right)}.
\label{calAQT}
\eeq
This is similar to the factor ${\cal A}(Q_T)$ of Ref.\ \cite{Boer-00}, but 
with the replacement of $J_0(b Q_T) \to J_2(b Q_T)$ 
in the numerator. We note that unlike the case of $A_{TT}^V$ investigated in 
Ref.\ \cite{Boer-00}, the present 
asymmetry does not need to oscillate as a function of $Q_T$. Rather the
vanishing of the numerator after integration over $d^2 \bm{q}_T^{}$ is due to
the angular integration. Note that 
$\kappa$ only depends on $Q_T$ and not on $\phi_1$. 
Also, we note that this asymmetry has a 
kinematic zero at $Q_T=0$, since $\bm{\hat h}$ 
cannot be defined in that case (and indeed $J_2(Q_T=0) = 0$). 
This is also seen from Ref.\
\cite{Boer-99}, where a similar asymmetry factor has been investigated as a
function of $Q_T$. 

Here the main focus will
be on the factor ${\cal A}(Q_T)$, which is a measure for the effect of the
Sudakov factors on the asymmetry compared to the tree level result
$\exp\left[-(R^2-R_u^2)Q_T^2/2\right] 
M^2 Q_T^2 R^6/R_u^2$ (cf.\ Eq.\ (\ref{treelevelA})), 
which is valid only for values of $Q_T$ of the 
order of the intrinsic transverse momentum. 

In the above expression we have introduced the usual cut-off $b_{\max}$, 
via $b_*=b/\sqrt{1+b^2/b_{\max}^2}$ and replaced $\frac{1}{2} \,
b^2/R^2 \to S_{NP}(b)$, for which we take one of the 
standard nonperturbative smearing functions, needed to describe the low 
$\bm{q}_T$ region properly. It is important to realize that $S_{NP}(b)$ is 
introduced only in part to take care of the smearing due to
the intrinsic transverse momentum (cf.\ Eq.\ (\ref{genNPform})), hence one
cannot simply equate $\frac{1}{2} \, b^2/R^2$ with $S_{NP}(b)$. 
But taking into account the term 
$\exp\left(-\frac{1}{2} \, b^2/R^2 \right)$ in addition to $S_{NP}(b)$ will 
just produce a change in the coefficient of the $b^2$ term in $S_{NP}(b)$. 
To keep the unpolarized cross section unaffected, we will therefore introduce
as nonperturbative term $\exp\left[{-S_{NP}(b)}
\, + \frac{1}{2}\,b^2/R^2 \right]$, in order not to count the contribution
from intrinsic transverse momentum twice. It is also worth mentioning again 
that since $R \neq R_u$ (important at tree level), $S_{NP}(b)$ need not be 
the same in numerator and denominator (much less relevant however, since it
affects only $g_A, g_B$ and not $g_Q$, cf.\ Eq.\ (\ref{genNPform})) 
and that in principle, it can depend 
on $z_1$ and $z_2$, but we will not take into account these refinements. 

\newpage 

Here we will take for the nonperturbative Sudakov factor 
the parameterization of 
Ladinsky-Yuan (Ref.\ \cite{Ladinsky-Yuan} with $ x_1 x_2 =10^{-2}$),
\beq
S_{NP}(b) = g_1b^2 + g_2 b^2
\ln\left(\frac{Q}{2Q_0}\right),
\label{LY94}
\eeq
with $g_1=0.11\,\text{GeV}^{2}, g_2=0.58\,\text{GeV}^{2}, 
Q_0=1.6 \,\text{GeV}$ and 
$b_{\max}= 0.5 \,\text{GeV}^{-1}$. 
Part of the results will depend considerably on this choice and this issue 
will be addressed in detail below. The
value $g_1=0.11\,\text{GeV}^{2}$ can be viewed as an intrinsic transverse
momentum of $\amp{\bm{p}_T^2} = 1/R^2 = (220\, \text{MeV})^2$. 

The reason we have chosen the parameterization of 
Ladinsky-Yuan \cite{Ladinsky-Yuan}, which is fitted to the transverse momentum
distribution of $W/Z$ production in $p p$  $(p \bar{p})$ scattering, is that
unfortunately there is no nonperturbative Sudakov factor available for $e^+
e^- \to A + B + X$, except for the  
energy-energy correlation function obtain from low energy data 
\cite{CS-PRL82}. This is surprising considering the wealth of data from LEP 
experiments. Moreover, the
nonperturbative Sudakov factor in the  
energy-energy correlation function ($\propto 
\sum_{A,B} \int dz_1 z_1 dz_2 z_2 Q^2 d\sigma/dQ_T^2$) as fitted in Ref.\ 
\cite{CS-PRL82} is not the same as the one in the differential cross section
as a function of the lightcone momentum fractions. For related discussions,
also relevant for SIDIS, see Refs.\ \cite{Meng-95,Nadolsky}.

Although the nonperturbative Sudakov factors for Drell-Yan and $e^+ e^-$ need 
not be related, the $S_{NP}$ of Ref.\ \cite{Ladinsky-Yuan} can be viewed
as a generic one and allows us to study the general features of the Sudakov
suppression. We will also investigate the dependence on the nonperturbative
Sudakov factor by varying the parameters. At a later stage one can always 
insert a more appropriate (phenomenologically determined) nonperturbative
Sudakov factor into the asymmetry expression Eq.\ (\ref{calAQT}). Our results
also underline the importance of a good determination of $S_{NP}$. 

\begin{figure}[htb]
\begin{center}
\leavevmode \epsfxsize=8cm \epsfbox{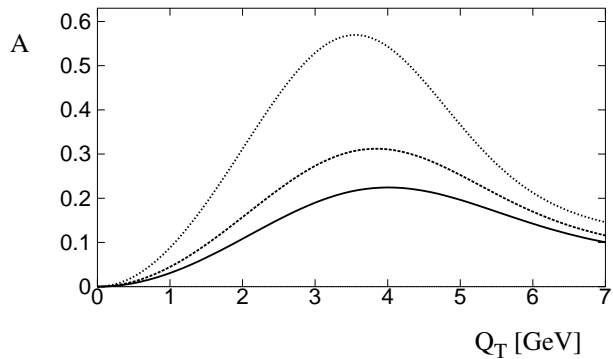}
\vspace{2 mm}
\caption{\label{Befr90}The asymmetry factor ${\cal A}(Q_T)$ (in units of
$M^2$) at $Q=30 \, \text{GeV}$ (upper curve), $Q = 60 \, \text{GeV}$ 
(middle curve) and at $Q = 90 \, \text{GeV}$.}
\end{center}
\vspace{-2 mm}
\end{figure}

In Fig.\ \ref{Befr90} 
the asymmetry factor ${\cal A}(Q_T)$ is given at the scales 
$Q=30 \, \text{GeV}$, $Q = 60 \, \text{GeV}$ and $Q=M_Z$. 
The maximum of the asymmetry factor ${{\cal A}(Q_T)}$ at $Q=30 \, \text{GeV}$ 
($S_{NP}(b) = 1.41 \, b^2$),  $Q = 60 \, \text{GeV}$ 
($S_{NP}(b) = 1.81 \, b^2$) and $Q=M_Z$ ($S_{NP}(b) = 2.05 \, b^2$) 
is seen to be $0.57$ (at $Q_T \sim 3.6 \, \text{GeV}$), 
$0.31$ (at $Q_T \sim 3.8 \, \text{GeV}$) and $0.22$ 
(at $Q_T \sim 4 \, \text{GeV}$), respectively. 
One observes that 
the magnitude of the asymmetry factor goes down with increasing 
energy and the position of the maximum --and also the
average $Q_T$-- shifts to higher values of $Q_T$. 

Now we will discuss the dependence of these results on the choice of $b_{\max}$
and $S_{NP}$. Taking a higher value of $b_{\max}$ increases the Gaussian 
width. The above choice of $b_{\max}= 0.5 \,\text{GeV}^{-1}$ can be 
considered as optimistic ($1/b_{\max}$ is the scale down to which one trusts 
perturbation theory), enhancing the asymmetry factor somewhat.   

\begin{figure}[htb]
\begin{center}
\leavevmode \epsfxsize=8cm \epsfbox{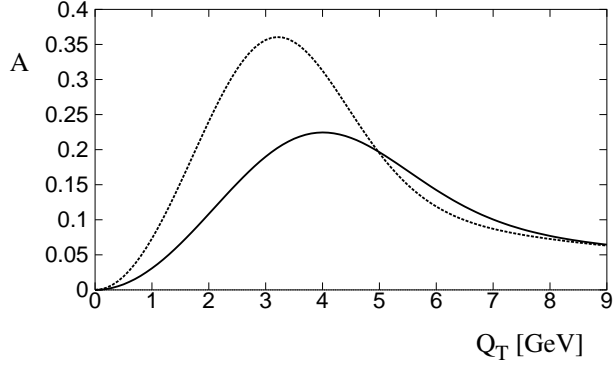}
\vspace{2 mm}
\caption{\label{Befr902}The asymmetry factor ${\cal A}(Q_T)$ (in units of
$M^2$) at $Q = 90 \, \text{GeV}$ with $S_{NP}(b) = 2.05 
\, b^2$ (solid curve) and $S_{NP}(b) = 1.37 \, b^2$ (dashed curve).}
\end{center}
\vspace{-2 mm}
\end{figure}

The asymmetry factor decreases with increasing 
Gaussian smearing width in $S_{NP}$, to which it has a considerable
sensitivity. Empirically, we find that if the
Gaussian width is reduced by a factor $\alpha$, then the maximum of the
asymmetry increases roughly by $\alpha$ and the corresponding value of $Q_T$ 
by $\sqrt{\alpha}$. In Fig.\ \ref{Befr902} this is illustrated for the 
asymmetry factor 
${{\cal A}(Q_T)}$ at $Q=M_Z$. The solid curve is for $S_{NP}(b) = 2.05 
\, b^2$ and the dashed curve for $S_{NP}(b) = 1.37 \, b^2$, where the latter
width is taken from a recent two parameter fit \cite{Landry}. 
The decrease with energy of the maximum of the asymmetry was found not to be 
very sensitive to changes to the Gaussian width in $S_{NP}$. 
We find that the decrease goes as $Q^{-0.9} - Q^{-1.0}$. 

So far we have only considered the asymmetry as a function of $Q_T$, which
means the cross section needs to be kept differential in the angle $\phi_1$ 
and the magnitude $Q_T$ of the transverse momentum $\bm{q}_T^{}$. Since the
asymmetry does not vanish after integration over $Q_T^2$ and since the 
$Q_T^2$-integrated cross section has been studied using LEP data 
\cite{Efremov}, we will now 
consider that case. Note that the integration has to be done
in numerator and denominator separately. Since the denominator has no
dependence on $\phi_1$ (the $\cos(2 \phi_1)$ dependence belongs to the 
numerator) one can in fact integrate over $\bm{q}_T^{}$
completely. This corresponds to the inclusive cross section in which
soft gluon contributions cancel, hence this integration over the
denominator of the asymmetry factor ${\cal A}(Q_T)$ yields 1 and has
no dependence on Sudakov factors. 

The $Q_T^2$-integrated result should be compared to tree level, 
therefore we will
first give the tree level expression. The tree level asymmetry is (neglecting
the small $c_3^a$ dependent term in the denominator) 
\beq
A^{(0)}(Q_T^{}) = \frac{Q_T^2 R^2 \exp(-R^2 Q_T^2/2)}{4 M^2 R_u^2 \exp(-R_u^2
Q_T^2/2)}\; \frac{\sin^2 \theta_2}{1 +
\cos^2\theta_2}\; 
\frac{\sum_{a}\;c_2^a\; H_1^{\perp a}(z_1) \; \overline H{}_1^{\perp a}(z_2)}{
\sum_{b}\; c_1^b\; D_1^b(z_1) \; \overline D{}_1^b(z_2)}. 
\label{treelevelA}
\eeq 
Note that if one integrates the 
cross section over $Q_T^2$, one should be careful to retain all dependences 
on $Q_T$ in both numerator and denominator separately. From the above
expression we also infer that ${\cal A}(Q_T)$ itself should be compared with 
the tree level quantity $\exp\left[-(R^2-R_u^2)Q_T^2/2\right] M^2 Q_T^2
R^6/R_u^2$ as mentioned before. 
\begin{figure}[htb]
\begin{center}
\leavevmode \epsfxsize=8cm \epsfbox{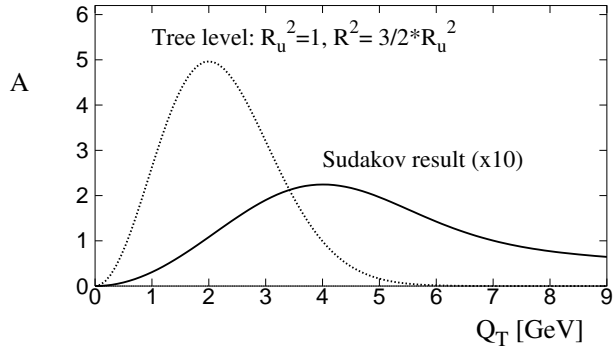}
\vspace{2 mm}
\caption{\label{Cos902}The asymmetry factor ${\cal A}(Q_T)$ (in units of
$M^2$) at $Q = 90 \, \text{GeV}$ multiplied by a factor 10 (solid
curve) and the tree level quantity (in units of
$M^2$) using $R_u^2=1 \, \text{GeV}^{-2}$ and
$R^2/R_u^2 = 3/2$.}  
\end{center}
\vspace{-2 mm}
\end{figure}
In Fig.\ \ref{Cos902} we have displayed the comparison of 
${\cal A}(Q_T)$ at $Q = 90 \, \text{GeV}$ and the tree level quantity 
using the values $R_u^2=1 \, \text{GeV}^{-2}$ and $R^2/R_u^2 = 3/2$, 
which were chosen such as to minimize the magnitude. The value 
$R_u^2=1 \, \text{GeV}^{-2}$ can be regarded as too small already. 
We conclude that inclusion of Sudakov factors has the effect of 
suppressing the tree level result by at least an order of magnitude. This is
important to keep in mind when making predictions of transverse momentum
dependent azimuthal spin asymmetries based on tree level
expressions. 

For the particular case of a produced $\pi^+$ and $\pi^-$, we will make the
following simplifying assumptions. We assume 
$D_1^{u\to\pi^+}(z)=D_1^{\bar d\to\pi^+}(z)$, 
$D_1^{d\to\pi^-}(z)=D_1^{\bar u\to\pi^-}(z)$ and neglect unfavored
fragmentation functions like $D^{d\to\pi^+}(z)$; and similarly
for the Collins functions. As a
consequence of these assumptions the fragmentation
functions can be taken outside the flavor summation and appear as a square
(at the average $z_1, z_2$). 
After $Q_T^2$ integration one arrives at a cross section differential in
the angle $\phi_1$ of $\bm{q}_T^{}$: 
\beq
\frac{d\sigma}{d\Omega d \phi_1} \propto 
\left( 1 + A\;\left[\frac{H_1^{\perp}}{D_1}\right]^2\;
F(y) \; \cos(2\phi_1)\right),
\label{myxs}
\eeq
where
\begin{equation}
F(y)=
\frac{\sum_{a=u,\bar d}\;c_1^e\,c_2^a\;B(y)}{\sum_{b=u,\bar d}\;
\left(c_1^e\,c_1^b\;A(y)-{\scriptstyle \frac{1}{2}}
\,c_3^e\,c_3^b\,C(y)\right)} \approx \frac{\sin^2 \theta_2}{1 +
\cos^2\theta_2}\; \frac{\sum_{a} \left(g_V^a{}^2 - g_A^a{}^2 \right)}{\sum_{b}
\left(g_V^b{}^2 + g_A^b{}^2 \right)},
\label{factor}
\end{equation}
which is largest at $\theta_2 =90^{\circ}$: $F_{\max} \approx -0.5$ 
(for a plot of the full factor $F(\theta_2)$ see Ref.\ \cite{BoerZ}). 

For the prefactor $A$ one finds at tree level $A^{(0)} = 1/(2 M^2 R^2) = 
\amp{\bm{p}_T^2}/(2 M^2)$ and one should assume a typical intrinsic 
transverse momentum squared value, in the range of 
$\amp{\bm{p}_T^2} \approx (200 - 700 \, \text{MeV})^2$. For pions this means 
$A^{(0)} \approx 1 - 12$. This can be compared to
$A^{(0)}=6/\pi$ of Ref.\ \cite{Efremov}, which yields $\amp{p_T^2} \approx 
(270 \, \text{MeV})^{2}$ (or $R^2 \approx 13 \, \text{GeV}^{-2}$), which seems
to be a reasonable value (remember that $R^2$ must be larger than $R_u^2$). 

After including the Sudakov factors our numerical calculation yields 
$A= 0.07$ ($S_{NP}(b, Q=M_Z) = 2.05 \, b^2$), much smaller than the tree
level values discussed above. The Gaussian 
width in $S_{NP}$ is not crucial for this
conclusion; if one takes a much smaller width of for instance 
$1 \, \text{GeV}^2$, then one obtains $A=0.11$. It has to be emphasized that 
the width used in $S_{NP}(b, Q)$ should increase with increasing energy. 

Our result shows that upon including Sudakov
factors one retrieves parton model characteristics (also noted in Ref.\
\cite{CS85e}), but with transverse momentum spreads that are significantly
larger than would be expected from intrinsic transverse momentum (this is
supported by the presently available parameterizations of $S_{NP}$ in various
processes, see e.g.\ \cite{Meng-95}). 
This means that a large tree level value for $A^{(0)}$ (e.g.\ $6/\pi$),
would lead to an extracted 
Collins function that is considerably smaller than if Sudakov 
factors would have been included. After the nonperturbative Sudakov factor 
has been obtained from LEP data, one can estimate exactly how much. For our
choice of $S_{NP}(b, Q=M_Z) = 2.05 \, b^2$, the effect on the magnitude of the
Collins function is an additional factor $\sqrt{6/(0.07 \pi)} \approx 5$ 
compared to the extraction using the tree level expression. 

We conclude that the Sudakov factors produce a
strong suppression compared to the tree level result. A tree level analysis 
applied at $Q=M_Z$ is expected to overestimate the asymmetry at least by an 
order of magnitude and therefore, it will underestimate the Collins 
function significantly. Hence, Sudakov factors should be taken into 
account when extracting the Collins function from LEP data or in general, from
$e^+ e^-$ data obtained at high values of $\sqrt{s}$. 

\section{Collins asymmetry in SIDIS}

The Collins function $H_1^\perp$ originally was shown 
to lead to a single spin azimuthal asymmetry in semi-inclusive DIS 
\cite{Collins-93b}. This
asymmetry has received much attention, since it would provide an additional
way of accessing the transversity function $h_1$. A preliminary determination 
of the Collins asymmetry from SMC
data has been performed \cite{Bravar-99}, yielding an asymmetry for $\pi^+$
production of $11\% \pm 6\%$. Also, the $\sin
\phi$ azimuthal asymmetry as recently measured by the HERMES Collaboration 
\cite{HERMES} might be related to the Collins asymmetry, providing further
indication that the Collins function is nonzero. Measurements by HERMES and 
COMPASS are expected to provide more conclusive information on 
the Collins effect and its magnitude. Since these
experiments are performed at different energies, it is important to know the
$Q^2$ dependence of the asymmetry. Here we will investigate the 
$Q^2$ dependence of the transverse momentum ($Q_T$) distribution of the 
asymmetry. 
 
From Ref.\ \cite{Boer-J-M-99} we extract the expression for this single spin
asymmetry (extended to include contributions from $Z$-boson exchange; the 
expressions for $W$-boson exchange can be found in Ref.\ \cite{Boer-J-M-99}): 
\beq
\frac{d\sigma(\ell H\to \ell' h X)}{dx dz dy d\phi_\ell d^{\,2}{\bm q_T^{}}}
= \frac{2\, \alpha^2\, x\, z^2\, s}{Q^4}\;\sum_{a,\bar a} \;\Bigg\{ 
          K_1^a(y)\;{\cal F}\left[f_1 D_1\right]
- |\bm S_{T}^{}|\;K_3^a(y)\sin(\phi_h^\ell+\phi_{S}^\ell)\;
             {\cal F}\left[\,\bm{\hat h}\!\cdot \!\bm{k}_T^{}\,
                    \frac{h_1H_1^{\perp}}{M_h}\right] + \ldots
\Bigg\} \;,
\eeq
where $\bm S_{T}^{}$ is the transverse spin of the incoming hadron $H$. 
The couplings $K_1^a(y)$ and $K_3^a(y)$ are of the same form as before, 
except that now
\ba
A(y) &=& \left(1-y+\frac{1}{2}y^2\right) \;, \\
B(y) &=& - (1-y) \;,\\[2 mm]
C(y) &=& - y (2-y) \;,
\ea
and $y= (P \cdot q)/ (P \cdot l) \approx q^-/l^-$ 
($l$ is the momentum of the beam 
lepton). 
Also, since $Q^2$ is now space-like, the width $\Gamma_Z$ can be ignored: 
\ba
\chi_{1} &=& \frac{1}{\sin^2 (2 \theta_W)} \, \frac{Q^2}{Q^2+M_Z^2} \;,\\
\chi_{2} &=&  \left(\chi_{1}\right)^2 \;. 
\ea 
The azimuthal angle $\phi_h^\ell$ ($\phi_S^\ell$) around the 
three-momentum of the virtual boson is between the lepton's 
three-momentum and the outgoing hadron's three-momentum (transverse spin),
cf.\ Ref.\ \cite{Boer-J-M-99}.
   
\begin{figure}[htb]
\begin{center}
\leavevmode \epsfxsize=8cm \epsfbox{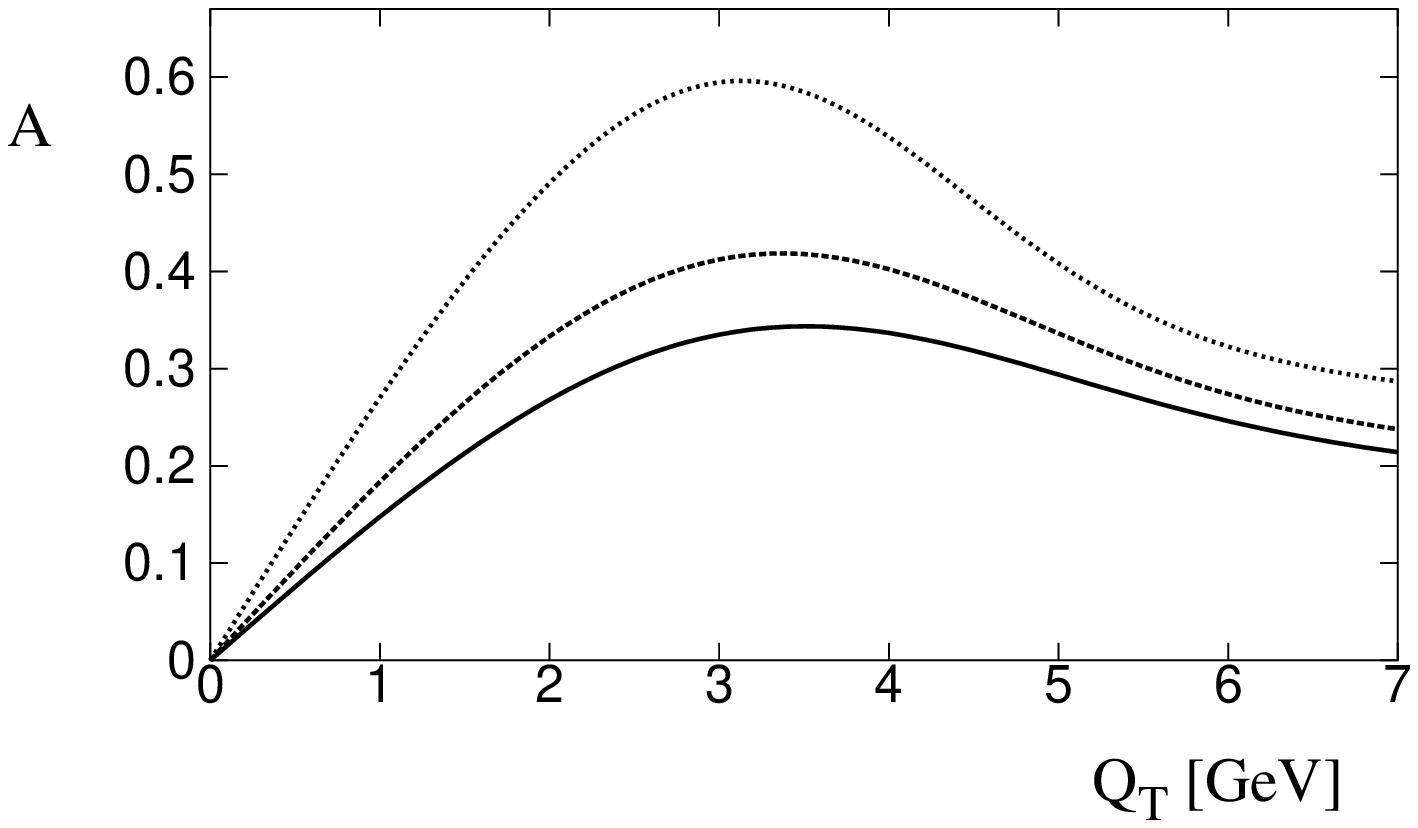}
\vspace{2 mm}
\caption{\label{Col90}The asymmetry factor ${\cal A}(Q_T)$ (in units of
$M_h$) at $Q=30 \, \text{GeV}$ (upper curve), $Q = 60 \, \text{GeV}$ 
(middle curve) and at $Q = 90 \, \text{GeV}$.}
\end{center}
\vspace{-2 mm}
\end{figure}

If we write the cross section as 
\beq
\frac{d\sigma(\ell H\to \ell' h X)}{dx dz dy d\phi_\ell d^{\,2}{\bm q_T^{}}}
\propto 
\left\{ 1 - |\bm S_{T}^{}|\;\sin(\phi_h^\ell+\phi_{S}^\ell)\; 
A(\bm{q}_T^{}) \right\},
\eeq
the asymmetry analyzing power is given by 
\beq
A(\bm{q}_T^{}) \equiv \frac{
\sum_{a}K_3^a(y)\;{\cal F}\left[\,\bm{q}_T^{}\!\cdot \!\bm{k}_T^{}\,
h_1 H_1^{\perp}\right]}{Q_T M_h \sum_{a} K_1^a(y)\; 
{\cal F}\left[f_1 D_1\right]}.
\eeq
Assuming again Gaussian transverse momentum dependence and including Sudakov
factors one arrives at
\beq
A(\bm{q}_T^{}) = \frac{
\sum_{a}\;K_3^a(y)\;h_1^a(x) H_1^{\perp a}(z)
}{2 M_h^2 R^2 \sum_{b}\; K_1^b(y)\;f_1^b(x) D_1^b(z)} {\cal A}(Q_T), 
\eeq
where $R^2$ is the Gaussian width of $H_1^\perp$ and 
the asymmetry factor ${\cal A}(Q_T)$ is defined as 
\beq
{\cal A}(Q_T) \equiv M_h \, \frac{
\int_0^\infty db \, b^2 \, J_1(b Q_T) \,  
\exp\left({-S(b_*)}\, 
{-S_{NP}(b)}\right)}{\int_0^\infty db \, b \, J_0(b Q_T)\, 
\exp\left({-S(b_*)}\, {-S_{NP}(b)} \right)}.
\label{calAQT2}
\eeq
This should be compared with Eq.\ (\ref{calAQT}) of the previous section.
In Fig.\ \ref{Col90} 
the asymmetry factor ${\cal A}(Q_T)$ is given at the scales $Q=30 \,
\text{GeV}$, $Q = 60 \, \text{GeV}$ and $Q=M_Z$. 
The maximum of the asymmetry factor ${{\cal A}(Q_T)}$ at 
$Q=30 \, \text{GeV}$, $Q = 60 \,
\text{GeV}$ and $Q=M_Z$ is seen to be $0.60$ (at $Q_T \sim 3.1 \,
\text{GeV}$), $0.42$ (at $Q_T \sim 3.4 \, \text{GeV}$) and $0.34$ 
(at $Q_T \sim 3.5 \, \text{GeV}$), respectively. Again we note that 
the magnitude of the 
asymmetry factor goes down with increasing energy and the position of the 
maximum shifts to higher values of $Q_T$. 

As before we have used the nonperturbative Sudakov factor by Ladinsky and Yuan,
Eq.\ (\ref{LY94}), in the absence of a well-established $S_{NP}$ for SIDIS 
(cf.\ Ref.\ \cite{Meng-95,Nadolsky}). But the results obtained here 
can nevertheless be viewed as generic, only changes in the specific numbers
are expected. Like for the previous $\cos(2\phi)$ asymmetry, the asymmetry 
factor decreases with increasing Gaussian smearing width. 
Empirically we find that if the Gaussian width in $S_{NP}$ is reduced by a 
factor $\alpha$, then both the maximum of the
asymmetry factor and the corresponding value of $Q_T$ increase roughly by
$\sqrt{\alpha}$. However, the decrease with energy of the maximum of the 
asymmetry is not very sensitive to changes in this Gaussian width. We find 
that the decrease goes as $Q^{-0.5} - Q^{-0.6}$, which is considerably slower
than for the $\cos(2\phi)$ asymmetry. In the next section we will discuss the
comparison between different asymmetries in more detail.

We will now compare these results to the tree level expression for this 
Collins asymmetry. At tree level the convolution ${\cal F}$ 
(where $w$ denotes a weight function) is given by 
\begin{equation} 
{\cal F}\left[w\left(\bm{p}_T^{},\bm{k}_T^{}\right) f\, D\right]\equiv \;
\int d^{\,2}\bm{p}_T^{}\; d^{\,2}\bm{k}_T^{}\;
\delta^2 (\bm{p}_T^{}+\bm q_T^{}-\bm 
k_T^{}) \, w\left(\bm{p}_T^{},\bm{k}_T^{}\right)  f^a(x,\bm{p}_T^2) 
D^a(z,z^2\bm{k}_T^2) \;.
\end{equation}
If Gaussian transverse momentum distribution and 
fragmentation functions are assumed, one obtains at tree level 
\beq
A^{(0)}(Q_T^{}) = \frac{Q_T R^2 \exp(-R^2 Q_T^2/2)}{2 M_h R_u^2 \exp(-R_u^2
Q_T^2/2)}\; \frac{
\sum_{a}\;K_3^a(y)\;h_1^a(x) H_1^{\perp a}(z)
}{\sum_{b}\; K_1^b(y)\;f_1^b(x) D_1^b(z)}.  
\label{treelevelAcol}
\eeq 
Therefore, one should compare ${\cal A}(Q_T)$ with 
the tree level quantity $\exp\left[-(R^2-R_u^2)Q_T^2/2\right] 
M_h Q_T R^4/R_u^2$. 
\begin{figure}[htb]
\begin{center}
\leavevmode \epsfxsize=8cm \epsfbox{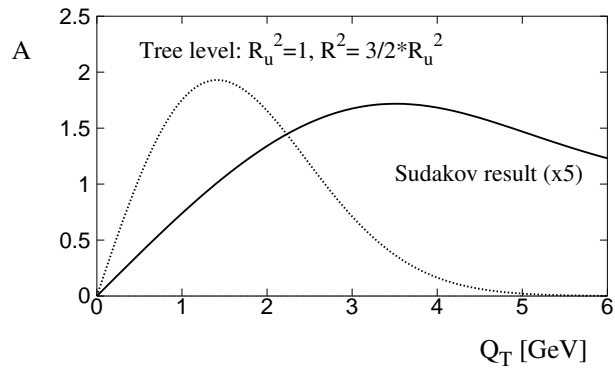}
\vspace{2 mm}
\caption{\label{Col902}The asymmetry factor ${\cal A}(Q_T)$ (in units of
$M_h$) at $Q = 90 \, \text{GeV}$ multiplied by a factor 5 (solid
curve) and the tree level quantity (in units of
$M_h$) using $R_u^2=1 \, \text{GeV}^{-2}$ and
$R^2/R_u^2 = 3/2$.}  
\end{center}
\vspace{-2 mm}
\end{figure}
In Fig.\ \ref{Col902} we have displayed the comparison of 
${\cal A}(Q_T)$ at $Q = 90 \, \text{GeV}$ and the tree level quantity using
again the values $R_u^2=1 \, \text{GeV}^{-2}$ and 
$R^2/R_u^2 = 3/2$, which minimize the maximum asymmetry value. 
As before we conclude that inclusion of Sudakov factors has the effect 
of suppressing the tree level result. It is also clear that for the Collins 
asymmetry (which arises from a single Collins effect) the Sudakov suppression 
is less severe than for the case of the $\cos(2 \phi)$ asymmetry 
(which depends on the Collins effect squared). 
One also observes a pronounced increase of the 
average $Q_T$ upon inclusion of Sudakov factors. The difference to tree level 
becomes even more pronounced as the 
choice of $R_u^2$ is increased to larger values more appropriate for a tree 
level analysis.  
Hence, also for transverse momentum dependent azimuthal spin asymmetries in
SIDIS one needs to include Sudakov factors in order to extract distribution
and fragmentation functions reliably.  

We now turn to a more general discussion of the different types of transverse
momentum dependent azimuthal spin asymmetries, categorized by different
transverse momentum weights and angular dependences. The comparisons are most
cleanly done for the asymptotic $Q^2$ case.   

\section{Asymptotic behavior}

In the previous sections we have studied asymmetries with 
different types of weights, namely  
$2\,\bm{\hat h}\!\cdot \! \bm p_T^{}\,\,\bm{\hat h}\!\cdot \! \bm k_T^{}\,
                    -\,\bm p_T^{}\!\cdot \! \bm k_T^{}$ 
and $\bm{\hat h}\!\cdot \! \bm k_T^{}$. 
The former (and also $\bm p_T^{}\!\cdot \! \bm k_T^{}$ in the asymmetry
$A_{TT}^V$ of Ref.\ \cite{Boer-00}) is typical of a double transverse spin 
asymmetry. It has
two powers of transverse momentum and is therefore expected to decrease faster
than the (single spin) asymmetries which have weights with a single power. 
The reason for this difference in decrease is
that the asymmetries, which are convolutions of transverse momentum 
distributions, are largest if these distributions are large and have a large 
overlap. The distributions are largest at small $\bm p_T^{}$ and 
$\bm k_T^{}$ and the overlap is largest at small transverse momentum 
$\bm{q}_T$. However,  
the explicit powers of transverse momentum in the weights tend to suppress the 
region of small transverse momentum. More powers of transverse momentum in the
weight thus implies more suppression. The power $n$ of $b^n$
in the numerator compared to the denominator is a measure of this effect. 
A weight
proportional to $\bm p_T^{i} \bm k_T^{j}$ will lead to $n=2$ (cf.\ Eq.\
(\ref{calAQT})), whereas $\bm
p_T^{i}$ or $\bm k_T^{j}$ will lead to $n=1$ (cf.\ Eq.\
(\ref{calAQT2})). Larger $n$ means, larger
suppression and since the transverse momentum distributions broaden and 
decrease in magnitude with increasing energy, the convolution also decreases.
Asymmetries with $n=1$ will fall off slower with energy and the Sudakov 
suppression is less severe. This natural expectation is clearly observed 
in the numerical analysis of the examples investigated here and in Ref.\
\cite{Boer-00}.     

It is interesting to see what is the asymptotic behavior of the Sudakov
suppression. As is usually done, we will assume that at high values of $Q$ the
nonperturbative factor is irrelevant (true only asymptotically) and consider 
asymmetry factors of the form 
\beq
{\cal A}_{n, m}(Q_T) \equiv M^n \, \frac{
\int_0^\infty db^2 \, b^n \, J_m(b Q_T) \,  
\exp\left({-S(b)}\right)}{\int_0^\infty db^2 \, J_0(b Q_T)\, 
\exp\left({-S(b)}\right)}.
\eeq
The asymptotic behavior of the denominator at $Q_T = 0$ (relevant for
the unpolarized differential cross section) has been studied before
\cite{Parisi-Petronzio,CSS85} by means of a saddle point approximation.  
Unfortunately, such a saddle point approximation cannot be applied to the
ratios ${\cal A}_{n, m}(Q_T)$ for $m\neq 0$ and $Q_T \neq 0$, since one 
cannot take $J_m(b Q_T) \approx J_m(b_{SP} Q_T)$ to deduce the $Q^2$ 
dependence. For instance, since $b_{SP} Q_T$ is in general not small, i.e.\ 
$\ll m$ (even under the assumption $Q_T^2 \ll Q^2$), no approximation to the
Bessel functions can be used in the region of interest (i.e., around 
the value of $Q_T$ where the asymmetry is maximal).

In Ref.\ \cite{Boer-00} the asymmetry factor ${\cal A}_{2, 0}(Q_T)$ arises, 
which does allow for a straightforward saddle point approximation at $Q_T=0$,
where the asymmetry is maximal. Therefore, we simply assume that the weight 
$2\,\bm{\hat h}\!\cdot \! \bm p_T^{}\,\,\bm{\hat h}\!\cdot \! \bm k_T^{}\,
                    -\,\bm p_T^{}\!\cdot \! \bm k_T^{}$
has similar behavior as $\bm p_T^{}\!\cdot \! \bm k_T^{}$. From our numerical
studies we conclude that the maximum of ${{\cal A}_{2, 2} (Q_T)}$ 
actually decreases faster with energy than the maximum of 
${{\cal A}_{2, 0} (Q_T)}$ (at $Q_T =0$). 
Moreover, for the asymptotic behavior of ${{\cal A}_{n, 0} (Q_T =0)}$ 
one can obtain an analytic expression. 
First we would like to comment on the saddle point approximation of 
expressions of the form 
\beq
\int_0^\infty db^2 \, b^n \, \exp\left({-S(b)}\right) =  \int
\frac{d\ln(b^2\Lambda^2)}{(\Lambda^2)^{1+\frac{n}{2}}} 
\exp \left\{ (1+\frac{n}{2})\ln(b^2\Lambda^2) - S(b) \right\}.
\label{integral}
\eeq
As before we will keep only the leading term in the
expansion of $A(\alpha_s(\mu))$ and $\alpha_s(\mu)/\pi= 1/(\beta_1
\ln(\mu^2/\Lambda^2))$:
\beq
S(b,Q)=\int_{b_0^2/b^2}^{Q^2} \frac{d \mu^2}{\mu^2} 
\frac{C_F}{\beta_1 \ln(\mu^2/\Lambda^2) } \, \ln \frac{Q^2}{{\mu}^2}. 
\eeq
The integral Eq.\ (\ref{integral}) has a saddle point at
\beq
b_{SP} = \frac{b_0}{\Lambda} \left( \frac{Q}{\Lambda}
\right)^{-C_F/[C_F+ (1+\frac{n}{2}) \beta_1]}. 
\eeq
For $n=0$ we retrieve the power $-0.41$ of Refs.\ 
\cite{Parisi-Petronzio,CSS85}. 
For general $n$ one finds in the saddle point 
approximation (defining $\gamma_n = - C_F/[C_F+ (1+\frac{n}{2}) \beta_1]$)
\beq
\int_0^\infty db^2 \, b^n \, \exp\left({-S(b)}\right)  \stackrel{SP}{\propto}
\left(\frac{b_0^2}{\Lambda^2}\right)^{(1+\frac{n}{2})} \; 
\left( \frac{Q^2}{\Lambda^2}
\right)^{(1+\frac{n}{2})\gamma_n + \frac{C_F}{\beta_1} 
(1+\gamma_n + \ln(-\gamma_n))}.
\eeq 
In this way one finds for the denominator of ${{\cal A}_{2, 0} (Q_T=0)}$
an approximate power behavior of $(Q^2)^{-0.94}$ and for the numerator
$(Q^2)^{-0.62}$, when one takes $C_F=4/3$ and $\beta_1 = 23/12$ 
(corresponding to 5 flavors, but 6 flavors only makes a few percent
difference). The approximated asymmetry 
factor ${{\cal A}_{2, 0} (Q_T=0)}$ then has an asymptotic power behavior of 
$(Q^2)^{-0.32}$, which is a similar decrease as found numerically in Ref.\
\cite{Boer-00} and apparently holds also for very large values of $Q$. 
At lower energies, the saddle point is not very pronounced and 
the introduction of the Gaussian smearing and $b_*$ will
have an effect on the approximation, hence we view this agreement as a
coincidence. Of course, the saddle point approximation should become better 
with increasing $Q^2$. 

For the ratios ${{\cal A}_{n, 0} (Q_T =0)}$ one can also derive an
analytic expression for the asymptotic $Q^2$ dependence:
\beq
{\cal A}_{n, 0}(Q_T=0) = M^n \, \frac{
\int_0^\infty db^2 \, b^n \,  \exp\left({-S(b)}\right)}{\int_0^\infty db^2 \, 
\exp\left({-S(b)}\right)} =  c_n \; 
\left(\frac{M^2 b_0^2}{\Lambda^2}\right)^{\frac{n}{2}}
\;\left( \frac{Q^2}{\Lambda^2} \right)^{\frac{C_F}{\beta_1} \ln c_n},
\eeq
where $c_n = (1 + C_F/\beta_1)/(1 + n/2 + C_F/\beta_1)$. For $n=2$ this yields
$(Q^2)^{-0.32}$ for 5 flavors, which shows that the saddle point approximation
is an extremely good approximation (the power of $Q^2$ differs by 
${\cal O}(10^{-9})$). One now expects that the maximum value of 
${{\cal A}_{1, 1} (Q_T)}$ 
(appearing in the Collins asymmetry) to fall off at least as fast as  
${{\cal A}_{1, 0} (Q_T =0)}$ (for all $Q$), which asymptotically goes 
as $(Q^2)^{-0.18}$. We note that the introduction of 
the Gaussian smearing and $b_*$ have a considerable effect on the 
approximation at lower energies as our numerical studies demonstrate
(in the studied range of $10 - 100 \, \text{GeV}$ 
we found approximately $1/Q$ for 
$\max({\cal A}_{2, 2} (Q_T))$ and $1/\sqrt{Q}$ for 
$\max({\cal A}_{1, 1} (Q_T))$).  
Our main conclusion is that for very high $Q$ transverse momentum 
dependent azimuthal spin asymmetries will fall off as a fractional power of
$1/Q$ and that the behavior (or an upper bound to it) can approximately be 
found by looking at the power of $b$ in the integrand. 

Finally, we want to mention that 
the increase of the Sudakov suppression with energy provides a solution to the
following problem. In general, azimuthal spin asymmetries in collinear 
configurations, where partons are collinear to the parent hadrons, are 
suppressed by explicit powers of the hard energy scales. The transverse 
momentum factors in the weights have to be generated in the hard scattering
part 
and for this one pays a price in terms of inverse powers of the hard scale. 
The collinear case can be viewed as averaged over the transverse
momenta, relevant for $\bm{q}_T^{}$-integrated cross sections, which receives
its main contribution from the region of small transverse momentum. 
On the other hand, in the case of $\bm{q}_T^{}$-dependent differential cross 
sections at small $\bm{q}_T^{}$ (compared to $Q$), 
small parton transverse momentum needs to be included (one deals with nearly 
collinear partons therefore). In this case there need not be an explicit 
power suppression to generate similar azimuthal spin asymmetries (the average 
transverse momentum is now a scale in the problem, so dimensionless ratios
using $Q_T$ rather than $Q$ can be formed). This seems counterintuitive, since
by considering a differential cross section for a nearly collinear
configuration, one can obtain information without power suppression, that is
power suppressed in the $\bm{q}_T^{}$-integrated cross section, which receives
its main contribution from the region of small transverse momentum also 
(for an example, see Ref.\ \cite{Boer-Japan}). But we find that instead 
of explicit power suppression the presence of Sudakov factors gives rise to 
partial power suppression. In this sense the power suppression is effectively 
replaced by a Sudakov suppression. This means that the $\bm{q}_T^{}$-dependent 
azimuthal spin asymmetries vanish with increasing energy, as do their
$\bm{q}_T^{}$-integrated counterparts. Hence, departing from the 
collinear configuration (that is, by including small
transverse momenta) does not allow one to measure azimuthal spin asymmetries 
in (differential) cross sections at arbitrary high energies.  

\section{Conclusions}

In this article we have shown by quantitative examples how transverse momentum 
dependent azimuthal spin asymmetries are suppressed by Sudakov 
factors, in the region where the transverse momentum is much smaller than the 
large energy scale $Q^2$. Physically, the Sudakov suppression stems from 
broadening of the transverse
momentum distribution due to recoil from soft gluon radiation and the
suppression increases with energy. This implies that tree level 
estimates of transverse momentum 
dependent azimuthal spin asymmetries tend to overestimate the magnitudes and
increasingly so with rising energy. 

The size and $Q^2$ dependence of the Sudakov suppression have 
been studied numerically for two such asymmetries, both arising due to the 
Collins effect. The size of the suppression (compared to tree level) 
depends considerably on the
nonperturbative Sudakov factor that must be determined from experiment;
however, the $Q^2$ dependence of the suppression turned out to be much less 
sensitive to the nonperturbative input. 

We observe that in general the larger the power $n$ of the transverse momentum
in the weight of an asymmetry, the larger the suppression. 
For the Collins effect driven $\cos(2 \phi)$ 
asymmetry in electron-positron annihilation into two almost back-to-back pions
($n=2$), 
the Sudakov suppression was numerically found to be approximately $1/Q$ for 
the maximum of the asymmetry and an upper bound for the asymptotic behavior 
was found to be $1/Q^{0.6}$. For the Collins effect single spin asymmetry
in semi-inclusive deep inelastic scattering ($n=1$), the Sudakov suppression 
was numerically found to be approximately $1/\sqrt{Q}$ for the maximum of
the asymmetry and an upper bound for the asymptotic behavior was found to be
$1/Q^{0.4}$. For the maximum of the asymmetries of the type 
${\cal A}_{n,0}(Q_T)$, the asymptotic $Q^2$ dependence could be calculated 
analytically. This provides upper bounds on the fall-off of 
${\cal A}_{n,m}(Q_T)$, with $m\neq 0$.

Since our results depend on the input for the 
nonperturbative Sudakov factors $S_{NP}$, which is not (well) determined for 
the above processes, the numerical conclusions 
about the size and $Q^2$ dependence of the suppression should be viewed 
as generic, not as precise predictions. Therefore, we would like to 
stress the need for an extraction of the nonperturbative Sudakov factor 
from the process $e^+ e^- \to A + B + X$ (for any two, almost back-to-back 
hadrons $A$ and $B$) and from SIDIS. Considering the wealth of data from the 
LEP and HERA experiments this should pose no problem. 

The Sudakov suppression of the transverse momentum distribution of
azimuthal spin asymmetries is significant already for $Q$ values in the range
of $10 - 100 \, \text{GeV}$ as can be concluded from comparison to tree 
level. In single spin asymmetries the magnitude of the suppression is less 
severe than in double spin asymmetries and in both cases a pronounced 
shift of 
the average $Q_T$ to higher values is observed. We conclude that it is
essential to take into account Sudakov factors in 
transverse momentum dependent azimuthal spin asymmetries.  

\acknowledgments 
I would like to thank John Collins, Anatoli Efremov, Zi-wei Lin, 
George Sterman and Werner Vogelsang 
for helpful comments and discussions. Also, I thank Chris Dawson for 
C$^{_{++}}$ help.
Furthermore, I thank RIKEN, Brookhaven National Laboratory and the U.S.\ 
Department of Energy (contract number DE-AC02-98CH10886) for
providing the facilities essential for the completion of this work.


\begin{thebibliography}{30}

\bibitem{Collins-93b}
J.C. Collins, Nucl.~Phys.~B 396 (1993) 161.

\bibitem{Boer} 
D. Boer, R. Jakob and P.J. Mulders, Nucl.~Phys.~B 504 (1997) 345.

\bibitem{BoerZ}
D. Boer, R. Jakob and P.J. Mulders, Phys.~Lett.~B 424 (1998) 143.

\bibitem{Boer-00} 
D. Boer, Phys.~Rev.~D 62 (2000) 094029.

\bibitem{CS81}
J.C. Collins and D.E. Soper, Nucl.~Phys.~B 193 (1981) 381; Nucl.~Phys.~B 213 
(1983) 545 (E).

\bibitem{CS85e}
J.C. Collins and D.E. Soper, Acta Phys.~Pol.~B 16 (1985) 1047.

\bibitem{CSS85} 
J.C. Collins, D.E. Soper and G. Sterman, Nucl.~Phys.~B 250 (1985) 199.

\bibitem{Jaffe-96}
R.L. Jaffe, hep-ph/9602236.

\bibitem{CSS83} 
J.C. Collins, D.E. Soper and G. Sterman, Phys.~Lett.~B 109 (1982) 388;
Nucl.~Phys.~B 223 (1983) 381; Phys.~Lett.~B 126 (1983) 275.

\bibitem{CSS-PLB84}
J.C. Collins, D.E. Soper and G. Sterman, Phys.~Lett.~B 134 (1984) 263.

\bibitem{Davies-Stirling}
C. Davies and W.J. Stirling, Nucl.~Phys.~B 244 (1984) 337.

\bibitem{Weber}
A. Weber, Nucl.~Phys.~B 382, 63 (1992); Nucl.~Phys.~B 403 (1993) 545.

\bibitem{Frixione}
S. Frixione, P. Nason and G. Ridolfi, Nucl.~Phys.~B 542 (1999) 311.

\bibitem{Boer-Mulders-00}
D. Boer and P.J. Mulders, Nucl.~Phys.~B 569 (2000) 505.

\bibitem{Ralst-S-79}
J.P. Ralston and D.E. Soper, Nucl.~Phys.~B 152 (1979) 109.

\bibitem{Tangerman-Mulders-95a}
R.D. Tangerman and P.J. Mulders, Phys.~Rev.~D 51 (1995) 3357; hep-ph/9408305.
 
\bibitem{Collins-93a}
J.C. Collins, Nucl.~Phys.~B 394 (1993) 169. 

\bibitem{Coll-S-82} 
J.C. Collins and D.E. Soper, Nucl.~Phys.~B 194, 445 (1982).

\bibitem{Mulders-Tangerman-96}
P.J. Mulders and R.D. Tangerman, Nucl.~Phys.~B 461 (1996) 197;
Nucl.~Phys.~B 484 (1997) 538 (E).

\bibitem{Col-89}
J.C. Collins, in {\em Perturbative Quantum Chromodynamics}, 
Ed.\ A.H. Mueller (World Scientific, Singapore, 1989), p.\ 573.

\bibitem{Boglione-Leader}
M. Boglione and E. Leader, Phys.~Rev.~D 61 (2000) 114001.

\bibitem{Boer-99}
D. Boer, Phys.~Rev.~D 60 (1999) 014012.

\bibitem{Ladinsky-Yuan}
G.A. Ladinsky and C.-P. Yuan, Phys.~Rev.~D 50 (1994) R4239.

\bibitem{CS-PRL82}
J.C. Collins and D.E. Soper, Phys.~Rev.~Lett.~48 (1982) 655;
Nucl.~Phys.~B 284 (1987) 253.

\bibitem{Meng-95}
R. Meng, F.I. Olness, and D.E. Soper, Phys.~Rev.~D 54 (1996) 1919.

\bibitem{Nadolsky}
P. Nadolsky, D.R. Stump, C.-P. Yuan, Phys.~Rev.~D 61 (2000) 014003.

\bibitem{Landry}
F. Landry, R. Brock, G. Ladinsky and C.-P. Yuan, Phys.~Rev.~D 63 (2001) 013004.

\bibitem{Efremov}
A.V. Efremov, O.G. Smirnova and L.G. Tkachev, Nucl.~Phys.~B (Proc.~Suppl.) 79 
(1999) 554; Nucl.~Phys.~B (Proc.~Suppl.) 74 (1999) 49.

\bibitem{Bravar-99}
A. Bravar (for the SMC Collaboration), Nucl.~Phys.~B (Proc.~Suppl.) 79 
(1999) 520.

\bibitem{HERMES}
A. Airapetian {\it et al.}, HERMES Collaboration, Phys.~Rev.~Lett.~84 
(2000) 4047.

\bibitem{Boer-J-M-99} 
D. Boer, R. Jakob and P.J. Mulders, Nucl.~Phys.~B 564 (2000) 471.

\bibitem{Parisi-Petronzio} 
G. Parisi and R. Petronzio, Nucl.~Phys.~B 154 (1979) 427.

\bibitem{Boer-Japan}
D. Boer, hep-ph/9912311.

\end{thebibliography}
\end{document}